\documentclass[preprint,12pt]{elsarticle}

\usepackage{amssymb}
\usepackage{amsmath}
\usepackage[hidelinks]{hyperref}
\usepackage{subcaption}
\usepackage{siunitx}

\journal{MNE}

\begin{document}

\begin{frontmatter}



\title{Fabrication and Validation of a MEMS Electron Monochromator} 


\author{M.J. Adriaans}

\author{J.P. Hoogenboom}
\author{A. Mohammadi-Gheidari\corref{cor1}}
\ead{A.M.Gheidari@tudelft.nl}
\cortext[cor1]{Corresponding author}
\affiliation{organization={Imaging Physics, Delft University of Technology},
            country={Netherlands}}

\begin{abstract}
Electron monochromators are essential components for many applications in electron microscopy but are often large and complex to operate instruments. In prior work, we presented a concept for monochromation using the fringe fields around miniature electrostatic elements. This concept could be realized with a compact monochromator design that requires tuning of only 7 power supplies. 

In this manuscript, we present the fabrication using MEMS technology and experimental testing of such a miniaturized monochromator. The fabrication scheme involves Bosch deep reactive ion etching to shape lens and deflector electrodes and focused ion beam milling to create nanoscale apertures in a silicon nitride membrane. All elements are then coated, precisely aligned and, after stacking, wire bonded to connect to external power supplies. We use the assembled prototype to measure the electron-beam energy spread in a scanning electron microscope demonstrating the energy filtering capability of the device. These results highlight the possibilities for making miniaturized electron optics instruments using alignment and stacking of different MEMS fabricated elements, representing an important step towards the development of compact, lightweight, high performance electron-beam instrumentation.
\end{abstract}




\end{frontmatter}



\section{Introduction}
\label{sec: Introduction}
Micro- and nano-fabrication holds great promise for creating new components for Electron Microscopy (EM) that are difficult or impractical to make using conventional machining techniques. 
Examples include nanostructured gratings and related structures for laser-driven electron acceleration and electron-wavefront shaping \cite{chlouba2023coherent,r2014efficient,krielaart2018grating,grillo2014highly,tavabi2021experimental}, 
miniaturized multipole elements \cite{krysztof2021fabrication,mohammadi2024charged}, and source components for multi-beam scanning electron microscopy \cite{zonnevylle2014multi,van2006multibeam,mohammadi2023multi,5271608}. 
Microfabrication has also been applied to the redesign of electron-mirror correctors  \cite{hartel2003mirror} using microelectromechanical systems (MEMS) components \cite{maas2025electrostatic}.
By integrating beam-deflection and aberration-correction functions into compact modular structures, such implementations could reduce system complexity and relax constraints on the surrounding electron-optical system. A novel, miniaturized EM component, could thus consist of various mutually aligned and stacked MEMS elements.

Recently, we have proposed a design for an electron monochromator based on precise stacking and bonding of several microfabricated lens, deflector, and aperture elements.
This fully electrostatic architecture is intended to provide high-performance electron monochromation in a compact and modular unit while requiring the tuning of only a few power supplies during operation. This could provide an attractive alternative for conventional electron monochromators, which are typically large and complex and require tuning of many (>50) power supplies to correct mechanical and field aberrations and thus reach optimal performance.
Practical realization of this MEMS electron monochromator however requires accurate fabrication of the individual electrodes and apertures, precise passive alignment of the stacked components, reliable bonding, and electrical connection of the miniaturized electrodes to the external power supplies.

In this work, we first summarize the operating principle and overall architecture of our MEMS monochromator. We then describe the main fabrication and assembly challenges and present the methods developed to address these challenges. 
Finally, we present first experimental results obtained with the assembled prototype and demonstrate basic energy-resolving functionality.
\section{Device layout and fabrication requirements}
\label{sec: layout and requirements}
\begin{figure}
    \centering
    \begin{subfigure}{0.49\textwidth}
        \centering
        \includegraphics[height = .6\linewidth]
        {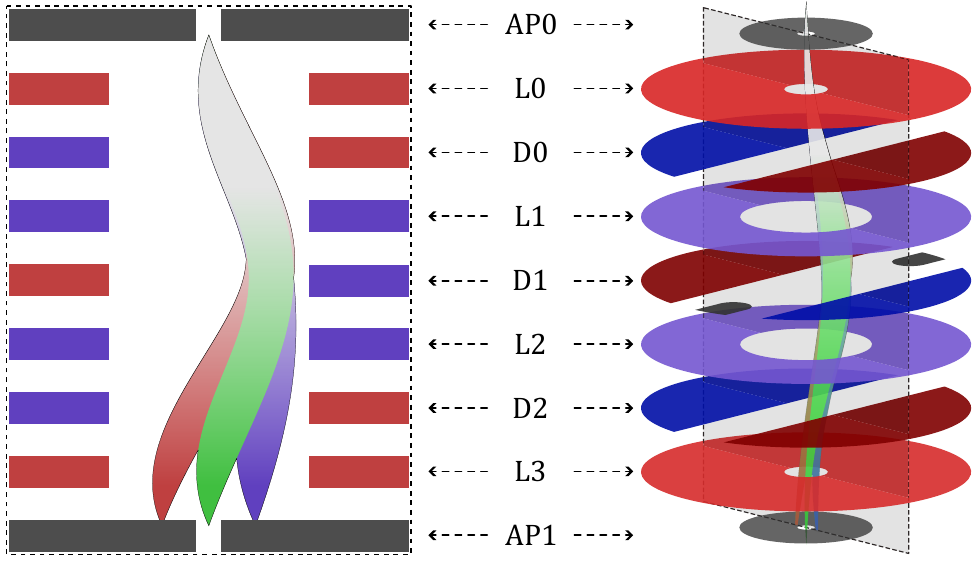}
        \caption{}
        \label{fig: mono layout a}
    \end{subfigure}
    \hfill
    \begin{subfigure}{0.49\textwidth}
        \centering
        \includegraphics[height = .6\linewidth]
        {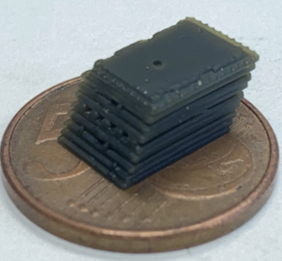}
        \caption{}
        \label{fig: mono layout b}
    \end{subfigure}
    \caption{
    (a) Schematic illustration of the electron-optical elements of the MEMS monochromator. The  left panel shows  a simplified 2D cross-section through the consecutive electrode layers, and the right panel shows a simplified 3D view of the electrodes stack. Red and blue indicate electrodes with  attracting and repelling biases respectively, while the aperture layers are shown in gray. The shades of red and blue are varied for visibility of the different shapes in the 3D view and are not intended to indicate a realistic proportional excitation. The stack comprises the rotationally symmetric lens electrodes L0–L3, the deflectors D0–D2, the entrance cleanup aperture AP0, and the energy-selection aperture AP1. Glass spacers define the separation between adjacent elements and provide electrical isolation. The unfiltered electron beam enters through AP0, is focused and energy-dispersed by the lens and deflector electrodes at AP1. The energy-selected portion of the beam is transmitted through AP1 and exits the monochromator.
    (b) 3D printed scale model of a MEMS Monochromator prototype on top of a 5 eurocents coin.
    }
    \label{fig: mono layout}
\end{figure}

The monochromator dimensions are largely determined by standard wafer thicknesses of the silicon wafers and glass spacers used in its construction. 
Before describing the fabrication process, we summarize the device layout presented in \cite{adriaans2026design} and the associated fabrication and assembly requirements. 
A schematic of the monochromator is shown in \autoref{fig: mono layout}.

The monochromator consists of a series of closely spaced conductive electrodes to which independently controlled voltages are applied to focus and deflect the electron beam. 
Glass spacers are used to define the axial distance between adjacent electrodes and deflectors and to electrically isolate them.
From entrance to exit, the stack is comprised of an initial beam cleanup aperture AP0, rotationally symmetric lens electrodes L0, L1, L2 and L3 and deflectors D0, D1 and D2, and an energy-selection aperture AP1. 
The electron beam enters through AP0 and is imaged and deflected by the lenses and deflectors towards AP1. The deflections cause the image to be spatially dispersed, in proportion to the energy spread around the nominal beam energy. This spatial separation of the different beam energies allows a narrow energy selection by AP1.
Because the lower half of the monochromator elements are constructed using rotated or laterally shifted versions of the upper half elements, only five distinct chip designs need to be fabricated: AP0, L0, D0, L1, and D1. 
The remaining elements L2, D2, L3, and AP1 are then produced using the corresponding designs from the first half of the stack. 
The electrode thicknesses are equal to standard silicon-wafer thicknesses, and most in-plane dimensions are several hundred micrometers, etched approximately parallel to the electron beam direction.
In contrast, AP0 and AP1 have nominal diameters of only 200 nm. 

The theoretical performance and tolerance analysis of this design have been described in detail in \cite{adriaans2026design}. 
The compact arrangement of the electrodes imposes several fabrication and assembly requirements.   
First, the electrode geometries should be fabricated with an accuracy of a few micrometers. 
Next, the relative alignment of successive  layers, as well as the alignment of the multipole electrodes within a layer should have an accuracy of within approximately 10 micrometers. 
Note that we keep this substantially smaller than the maximum allowed linear displacement of 53 micrometers calculated in \cite{adriaans2026design}, in which case the resolution would be substantially affected.
Additional critical requirements concern the size, shape, and quality of the nanometer-scale apertures and their alignment relative to the other electron-optical elements. 
Provided these requirements are met, fabrication and assembly errors are not expected to significantly degrade the predicted monochromator performance. 
Finally, the electrode surfaces should be conductive and electrically isolated from one another, with independent connections to the external power supplies.
\section{Fabrication and assembly}
\label{sec: fabrication}
\subsection{General electrode fabrication process}
The design requirements necessitate distinct fabrication strategies for the focusing or deflecting electrodes and the nano-apertures.
MEMS micromachining enables the electrode geometries to be defined with an accuracy of a few micrometers, while passive mechanical alignment can provide the required layer-to-layer positioning accuracy \cite{slocum2003precision,ling2000passive,boudreau2018passive}. 
Any spacer inserted in between two adjacent electrodes should also be able to withstand hundreds of volts applied between
the electrodes. 

One possible approach is to use glass rods or spheres that self-align within trenches etched into the individual elements. 
However, such an arrangement would concentrate the assembly forces at the point of contact between the glass elements and the sharply etched silicon edges. 
Although this configuration may be mechanically stable under nominal loading conditions, an inadvertently applied excessive force during assembly could fracture a sharply etched silicon edge, resulting in both misalignment and particle generation. 
Such damage might remain undetected until the completed monochromator is tested. We therefore use flat spacers laser-cut from borosilicate glass wafers. 
These spacers provide broad-area mechanical support, define the out-of-plane separation between adjacent elements, and provide electrical isolation. 
The in-plane position and rotation of each element are subsequently defined relative to an external mechanical reference. Accordingly, dedicated alignment features are incorporated into the design of each element (further adressed in \autoref{sec: stacking and wirebonding}).

\begin{figure}
    \centering
    \begin{subfigure}{0.49\textwidth}
        \centering
        \includegraphics[height=1\textwidth]{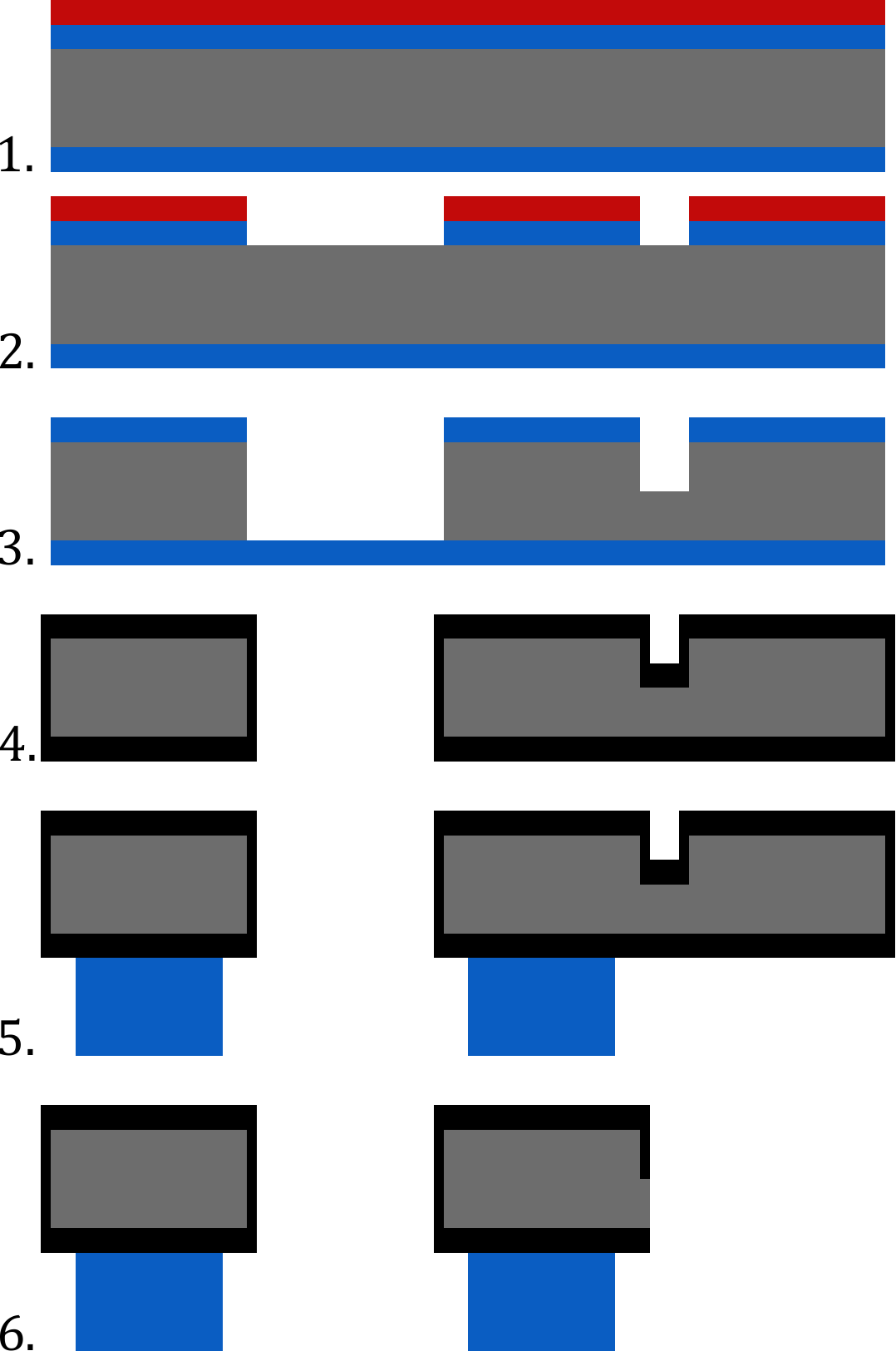}
        \caption{}
        \label{fig: basic fabrication steps a}
    \end{subfigure}
    \hfill
    \begin{subfigure}{0.49\textwidth}
        \centering
        \includegraphics[height=1\textwidth]{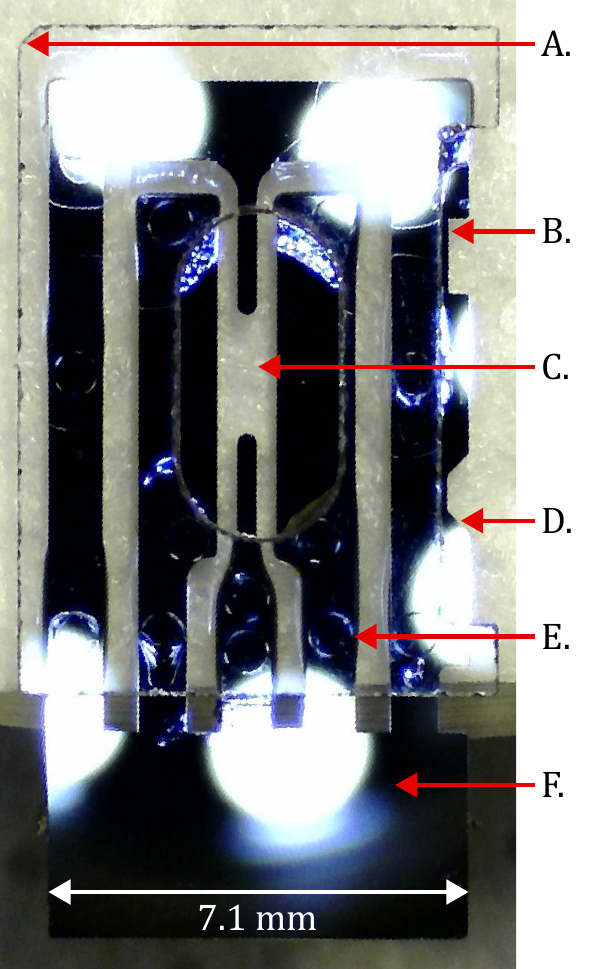}
        \caption{}
        \label{fig: basic fabrication steps b}
    \end{subfigure}
    \caption{Fabrication and assembly of the MEMS electrode elements. (a) Schematic process sequence: 1. PECVD deposition of $\textrm{SiO}_2$ (blue) and application of photoresist (red) 
    2. Photolithographic patterning of the $\textrm{SiO}_2$ hard mask. 
    3. Through-wafer Bosch deep reactive-ion etching. 
    4. Removal of the hard mask and sputter coating with Mo (black). 
    5. Adhesive bonding of the electrode chip to a laser-cut borosilicate-glass spacer (blue). 
    6. Fracture and removal of the retainer frame along the predefined break lines. 
    (b) Optical micrograph of a D1 multipole bonded to its glass spacer before removal of the retainer frame. The indicated features include A. a spacer-orientation feature, B. and D. U- respectively V-grooves for passive alignment, C. the central slot hole providing space for an unobstructed beam path and the 4 electrodes surrounding the beam path, E. the epoxy-collection trenches, F. retainer frame. The white spots are reflections.
    }
    \label{fig: basic fabrication steps}
\end{figure}

The general processes required to shape the deflecting and focusing electrodes are schematically shown in \autoref{fig: basic fabrication steps a}. 
Each electrode layer is patterned and etched in a single lithographic process. 
First, a $\textrm{SiO}_2$ layer is deposited by plasma-enhanced chemical vapor deposition (PECVD) and patterned photolithographically to form a hard mask with Reactive Ion Etching (RIE). 
The electrode geometry is then etched through the entire silicon wafer using Bosch deep reactive-ion etching. 
For the deflector and quadrupole layers, all constituent electrodes are formed simultaneously from the same wafer, thereby preserving their relative positions with micrometer-scale accuracy.

Each multipole-electrode pattern includes a retainer frame that temporarily holds the individual electrodes in position. 
The electrodes are connected to this frame through predefined break lines. 
Each break line is formed by a \qty{5}{\um} wide trench. 
Owing to the aspect-ratio-dependent etching rate, this narrow trench does not penetrate the full wafer thickness and therefore leaves a thin silicon connection that maintains the relative positions of the electrodes during subsequent processing. 
After removal of the hard mask by vapor etching, the silicon surfaces are sputter-coated with molybdenum (Mo) to provide electrical conductivity. 
The electrode chip is then bonded to a borosilicate-glass spacer. 
For the multipole layers, the retainer frame is subsequently fractured along the break lines and removed, leaving the individual electrodes fixed to the glass spacer.

As an example, \autoref{fig: basic fabrication steps b} shows a D1 multipole bonded to its spacer before removal of the retainer frame.
Several additional design features are visible in \autoref{fig: basic fabrication steps b}.
Over most of the chip perimeter, the glass spacer extends beyond the conductive silicon edges to reduce the risk of electrical discharge. 
At selected locations, U- and V-grooves, which act as our alignment features, extend beyond the perimeter of the glass.
The electrodes extend beyond a slotted hole in the center of the glass spacer, such that the fields that affect the passing electron beam are not affected by the glass spacer.
Moreover, the glass spacer extends up to the break line at the bottom of the chip, where the chips must be accessible for wire bonding later.

A UV-curing epoxy containing \qty{9}{\um} diameter glass microspheres is used for bonding. The microspheres establish a reproducible bond-line thickness between the silicon chip and the glass spacer and prevent smaller surface defects or contaminating particles from determining this separation. The deposited epoxy volume is selected to provide sufficient bonded area without overflowing onto critical conductive surfaces. In addition, concentric circular trenches, each \qty{5}{\um} wide, are  etched into the silicon chip to collect excess epoxy and prevent it from spreading into the electron-optically relevant  or wire-bonding regions. These epoxy-collection trenches are visible in \autoref{fig: basic fabrication steps b} (marked as E.).
\subsection{Nano-aperture fabrication}
\begin{figure}
    \centering
    \begin{subfigure}{.4\textwidth}
        \centering
        \includegraphics[height=0.8\textwidth]{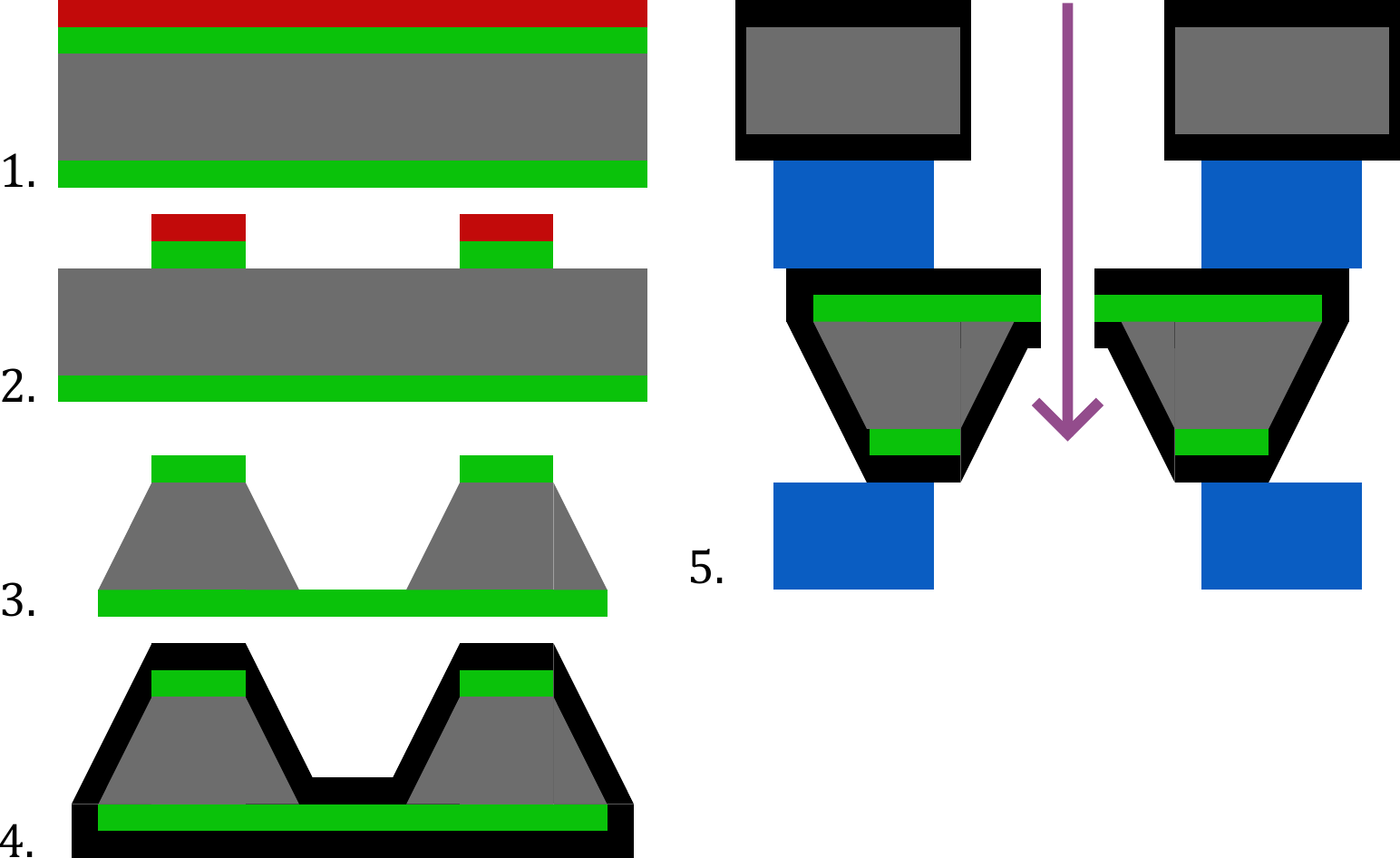}
        \caption{}
        \label{fig: membrane fabrication steps a}
    \end{subfigure}
    \hfill
    \begin{subfigure}{.4\textwidth}
        \centering
        \includegraphics[height=0.8\textwidth]{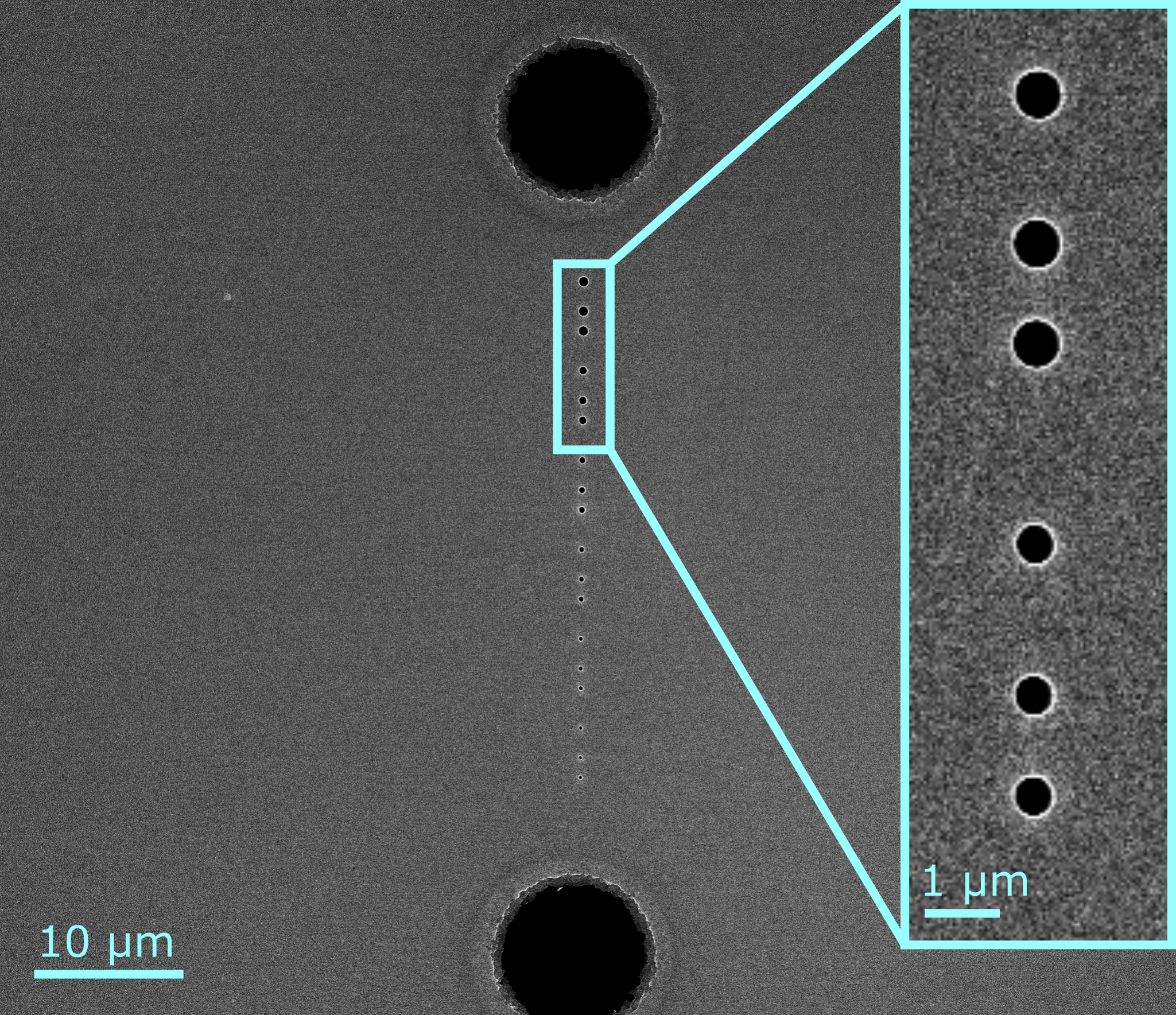}
        \caption{}
        \label{fig: membrane fabrication steps b}
    \end{subfigure}
    \caption{(a) Fabrication scheme for the entrance and selection apertures AP0 and AP1. 
    1. Deposition of $\textrm{Si}_3 \textrm{N}_4$ (green) and application of photoresist (red). 
    2. Patterning $\textrm{Si}_3 \textrm{N}_4$ mask.  
    3. Removal of photoresist and anisotropic wet etching of the silicon substrate in KOH. 
    4. Sputter coating with Mo (black). 
    5. Bonding of the aperture chip to the adjacent lens-electrode layer, followed by aligned focused-ion-beam milling of the nano-aperture. 
    (b) SEM micrograph of representative focused ion beam milling result. The \qty{10}{\um} apertures are for course testing and beam alignment, and the nano-apertures are for energy selection.
    }
    \label{fig: membrane fabrication steps}
\end{figure}

Due to their smaller size, AP0 and AP1 cannot be etched through a wafer using the same procedure. 
Instead, the AP0 and AP1 layers are fabricated as molybdenum-coated $\textrm{Si}_3 \textrm{N}_4$ membranes.
The fabrication process for these apertures is illustrated in \autoref{fig: membrane fabrication steps a}. 
A $\textrm{Si}_3 \textrm{N}_4$ layer is first deposited by Low Pressure Chemical Vapor Deposition (LPCVD) on a silicon substrate and patterned to form a hard mask with RIE. The underlying silicon is then anisotropically etched in KOH, leaving a free-standing $\textrm{Si}_3 \textrm{N}_4$ membrane. 
The membrane is subsequently sputter-coated with molybdenum to provide a conductive surface and prevent charging under electron-beam irradiation. 

In principle, the nano-apertures could be fabricated by electron-beam lithography followed by etching, either before or after releasing the membrane. 
However, the apertures have to be aligned with respect to the other monochromator elements. 
Nano apertures etched into the membrane would thus have to be aligned to the larger features etched from the other side. 
Furthermore, while KOH etching can produce a V-groove, the geometry and orientation is constrained by the crystallographic planes of the silicon substrate and cannot reproduce the required alignment feature in this design. 
Additional lithographic and etching steps would therefore be needed, unnecessarily increasing the complexity of the fabrication process. 
Thus, the membrane chip is instead first coarsely aligned and bonded to the adjacent conductive electrode layer. 
The apertures are subsequently milled by focused ion beam (FIB) using features on the neighboring electrode for alignment of the nano-apertures. 
A representative FIB-milling result is shown in \autoref{fig: membrane fabrication steps b}. 
The pattern comprises two coarse-alignment apertures with diameters of \qty{10}{\um} and an array of nano-apertures with diameters ranging from 200 to 600 nm for energy selection. 
Though the theoretical optical performance of the monochromator is based on 200 nm apertures, larger apertures are included since these are easier to focus a low beam energy on in our setup
Moreover, the apertures can close or get otherwise damaged by electron beam exposure which would take more time for bigger apertures.
Since even the bigger apertures could get clogged up, we require many apertures close to the optical axis as backup.
When focusing a beam with a broader spatial distribution on an aperture in a tightly packed regular array on AP0, a second or even more portions of the beam could pass through other accidentally aligned apertures on AP1, producing secondary beams. 
In order to make sure  only one nano-aperture pair aligns, the arrays on AP0 and AP1 are positioned such that they form a Moiré pattern when projected onto one another. 
This allows for small center-to-center distances between the apertures within each array while preventing neighboring apertures in the two layers from becoming mutually aligned. 
Consequently, only the intended aperture pair forms a continuous transmission path through the monochromator, thereby suppressing parasitic secondary beams passing through neighboring apertures.

\subsection{Stacking and wirebonding}
\label{sec: stacking and wirebonding}
\begin{figure}
    \centering
    \begin{subfigure}{0.49\textwidth}
        \centering
        \includegraphics[height=0.8\textwidth]{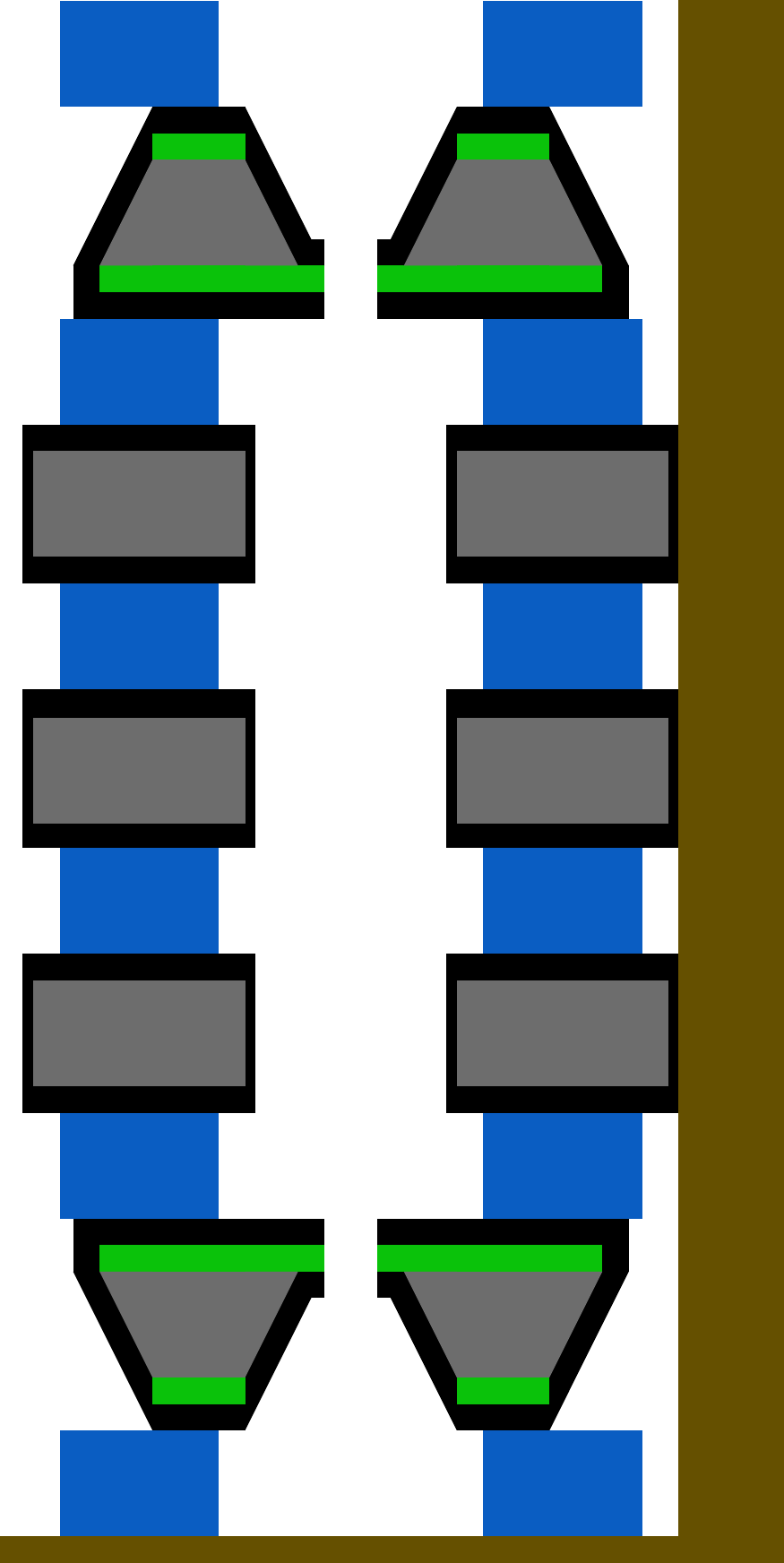}
        \caption{}
        \label{fig: stacking a}
    \end{subfigure}
    \hfill
    \begin{subfigure}{0.49\textwidth}
        \centering
        \includegraphics[height=0.8\textwidth]{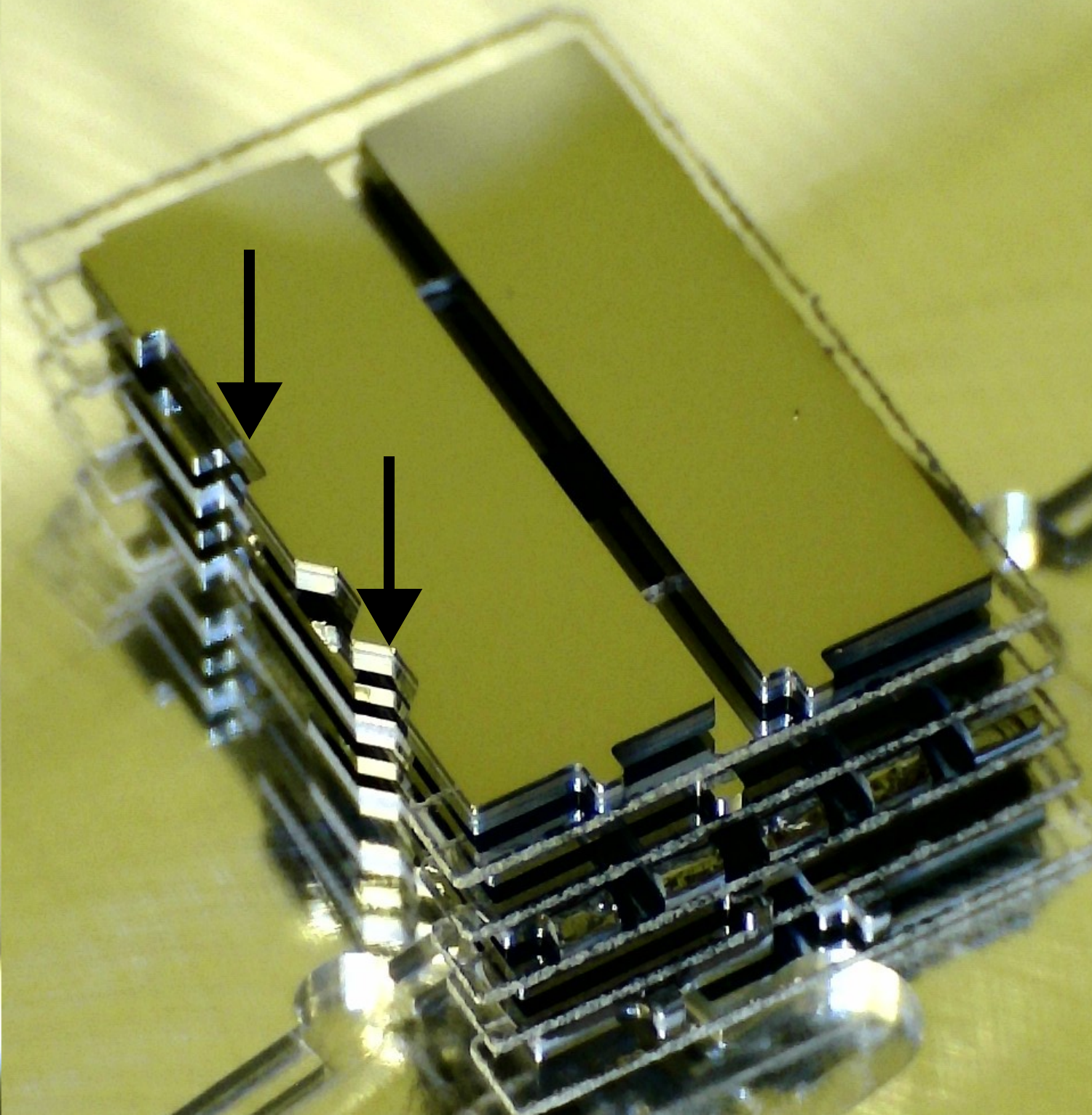}
        \caption{}
        \label{fig: stacking b}
    \end{subfigure}
    \caption{Stacking and passive alignment of the MEMS electrode layers. 
    (a) Two-dimensional schematic of the passive-alignment method. Alignment features on the electrode chips are brought into contact with two cylindrical reference pins in a mechanical jig.  
    (b) Optical micrograph of the aligned central subassembly extending from D0 to D2, before the addition of the entrance and exit subassemblies containing the aperture membranes. The arrows indicate the aligned U- and V-grooves. The width of each conductive element is 7.1 mm.}
    \label{fig: stacking}
\end{figure}

\begin{figure}
    \centering
    \begin{subfigure}{0.49\textwidth}
        \centering
        \includegraphics[height=0.8\textwidth]{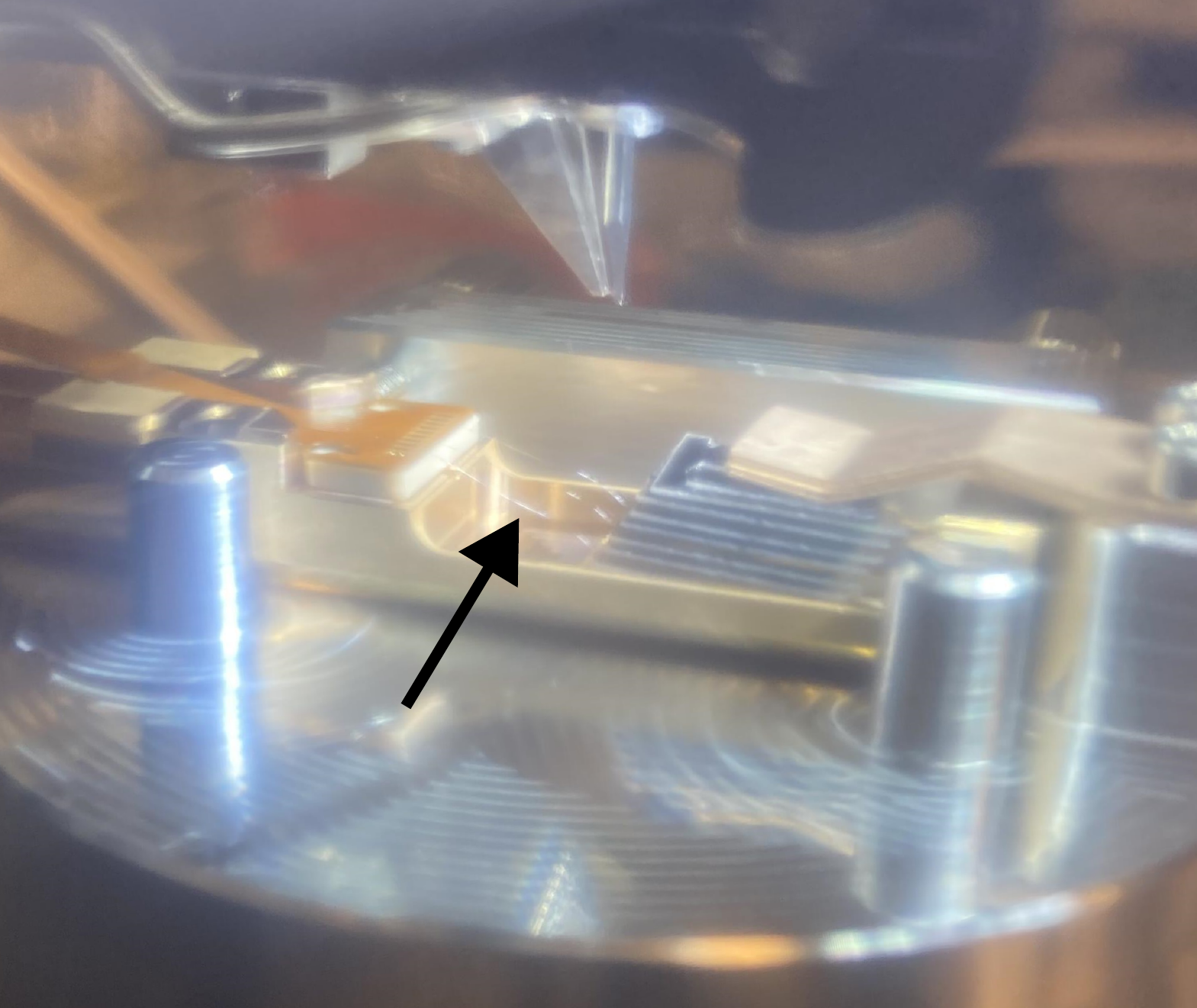}
        \caption{}
        \label{fig: wirebonding a}
    \end{subfigure}
    \hfill
    \begin{subfigure}{0.49\textwidth}
        \centering
        \includegraphics[height=0.8\textwidth]{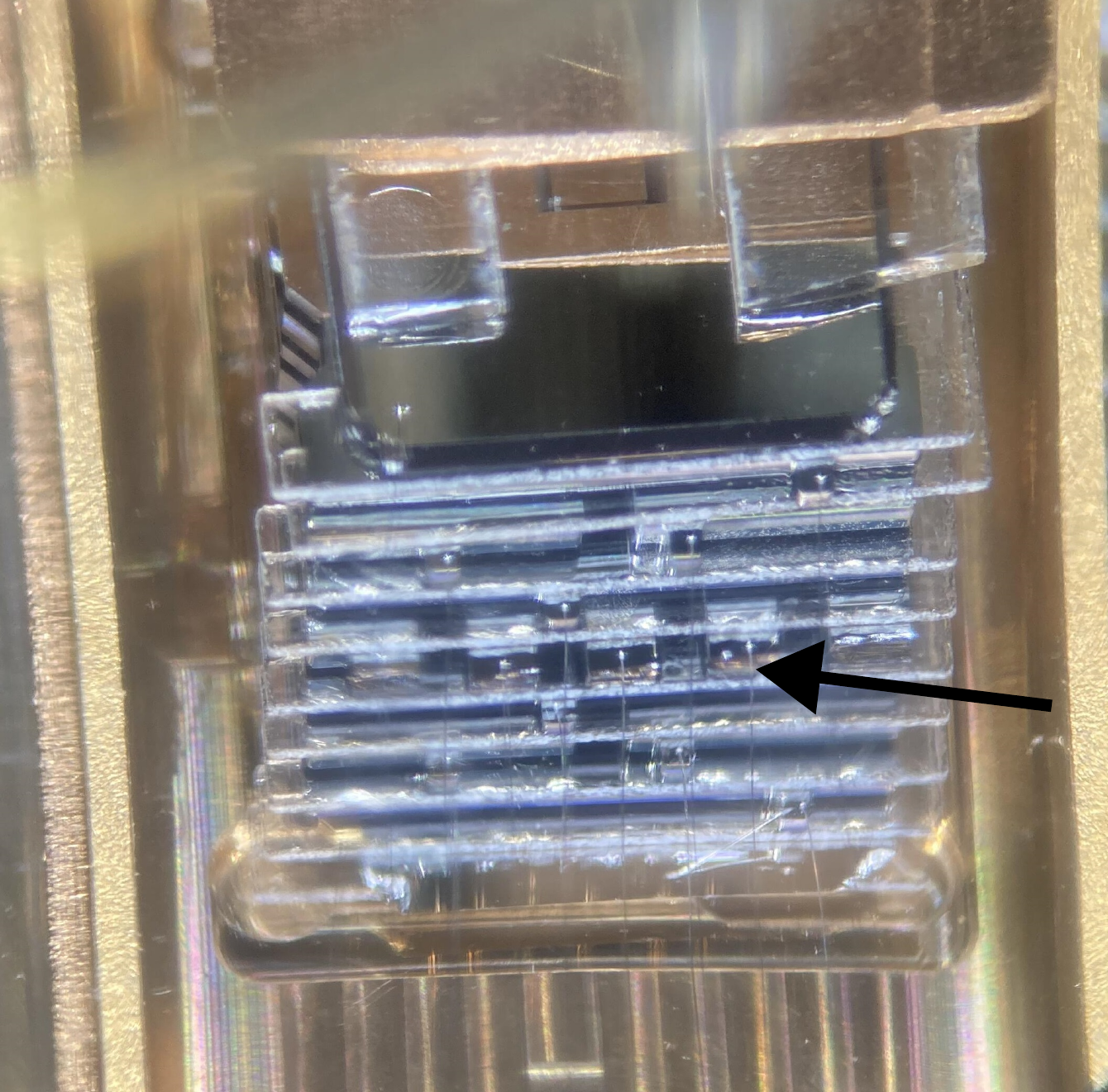}
        \caption{}
        \label{fig: wirebonding b}
    \end{subfigure}
    \caption{ Wire bonding of the assembled MEMS monochromator. 
    (a) Side view of the monochromator and wire-bonding arrangement in the assembly holder. 
    (b) Front view of the stack after completed wire bonding. In both panels, the black arrows indicate a long bonded wire connecting the electrode stack to the flex PCB. Bonded wires span more than 10 mm and reach loop heights of approximately 5 mm.}
    \label{fig: wirebonding}
\end{figure}

After fabrication and bonding of the individual electrodes to their glass spacers, the electrode layers need to be aligned and stacked.
For alignment, features etched into the sides of the electrodes are pushed against two reference cylinders in a mechanical assembly jig. 
To constrain the remaining three in-plane mechanical degrees of freedom, the features are a combination of a V-groove and a shallow U-groove such that the reference cylinders should touch the faces of these grooves at exactly three well-defined surfaces for each element.
This alignment procedure is illustrated schematically in \autoref{fig: stacking a}. 
The alignment features etched into the sides of the electrodes are pushed against two reference cylinders in a mechanical jig. 
These contact points constrain the in-plane position and rotation of each layer relative to a common mechanical reference.
The complete monochromator chip is assembled in stages:
First, a central subassembly extending  from D0 to D2 is stacked and aligned. 
This intermediate assembly  is shown in  \autoref{fig: stacking b}. 
The entrance and exit subassemblies, containing AP0 and L0 and L3 and AP1, respectively, are subsequently added to complete the stack. 
The assembled monochromator is then wire-bonded to a flexible printed circuit board (flex PCB), which provides independent electrical connections between the electrode layers and the external power supplies. 
The wire-bonding arrangement is shown from the side in \autoref{fig: wirebonding a} and from the front in \autoref{fig: wirebonding b}. 

To provide electrical access to each layer, the electrodes are laterally offset to form a staircase arrangement, such that the bonding pads of each layer remain exposed beyond the edge of the glass spacer above. In this prototype, the bonded wires indicated by the black arrows in \autoref{fig: wirebonding a} and \autoref{fig: wirebonding b} span distances of more than 10 mm between the flex PCB and the electrode stack and reach loop heights of approximately 5 mm.
The staircase pattern requires a slotted hole in each glass spacer allowing the electrodes to protrude beyond the spacers near the path of the electron beam while the in-plane position of each glass spacer varies along the height of the stack. 
The laser-cut glass spacers exhibited residual contamination and dimensional variations that made them difficult to clean and prevented a precise fit within the prototype holder. 
Consequently, the monochromator stack is slightly tilted in the holder in \autoref{fig: wirebonding a}
This tilt could be reduced by carefully repositioning the stack within the holder.

\section{Experimental characterization}
\label{sec: experimental results}
\subsection{Experimental setup and measurement method}
\begin{figure}
    \centering
    \begin{subfigure}{0.49\textwidth}
        \centering
        \includegraphics[height=0.9\textwidth]{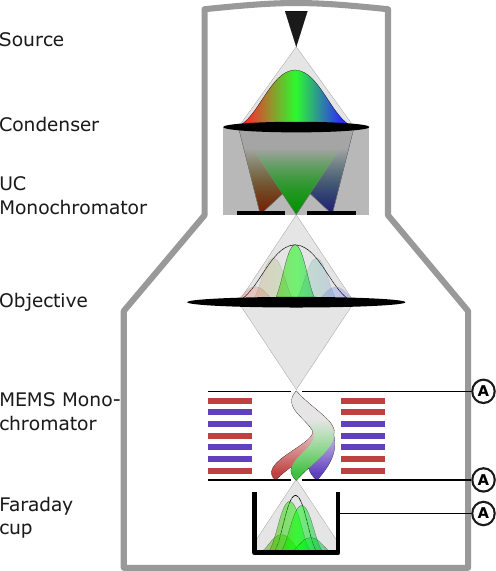}
        \caption{}
        \label{fig: experimental setup a}
    \end{subfigure}
    \hfill
    \begin{subfigure}{0.49\textwidth}
        \centering
        \includegraphics[height=0.9\textwidth]{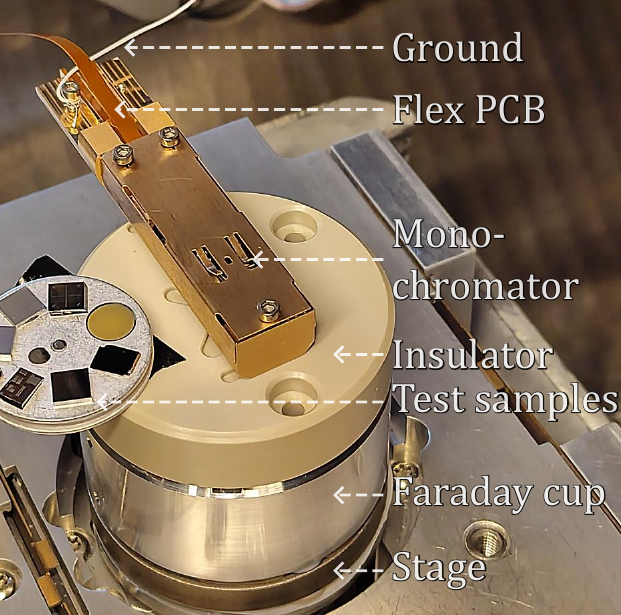}
        \caption{}
        \label{fig: experimental setup b}
    \end{subfigure}
    \caption{Experimental characterization of the MEMS monochromator. 
    (a) Schematic of the experimental arrangement in the SEM. The UC monochromator can be disabled to provide the unfiltered source-energy distribution or enabled to produce an incident beam with a reduced energy spread. The microscope objective optics focus the beam into the MEMS monochromator, which is operated as a scanning energy analyzer. The transmitted current is measured using a Faraday cup. The current-meter symbols indicate the planes at which the beam current can be monitored during alignment. 
    (b) Photograph of the MEMS monochromator, flex PCB, Faraday cup, and associated components mounted on the microscope stage.}
    \label{fig: experimental setup}
\end{figure}

The experimental configuration is shown schematically in \autoref{fig: experimental setup a}, and the MEMS monochromator mounted on the sample stage of the scanning electron microscope (SEM)  is shown in \autoref{fig: experimental setup b}.
The measurements were performed using a Thermo-Fisher Scientific Helios DualBeam system equipped with a UC monochromator integrated into the condenser-optics system. 
When enabled, the UC monochromator produces an electron beam with a reported energy spread of approximately 150 meV \cite{henstra2009versatile}. 
The MEMS monochromator is used to sample an energy-filtered or an unfiltered incident beam, by switching the UC monochromator on or off respectively.
First, the beam was focused onto one of the \qty{10}{\um} diameter coarse-alignment apertures in AP0. The MEMS monochromator was then aligned so that the energy-dispersed beam was focussed and dispersed at the corresponding coarse-alignment aperture in AP1. 
The nominal beam pass energy for the monochromator design is $E=\qty{500}{eV}$.
To vary the pass energy ($E$) of the MEMS monochromator by a given scaling factor, all applied electrode voltages are multiplied by the same common scaling factor, which was swept over a series of discrete values. For each value of the scaling factor, the transmitted current was recorded using the Faraday cup, thereby producing a transmission curve ($I(E)$). 
We expect the relatively large diameter of the coarse alignment aperture to be substantially larger than the spatial extent of the energy-dispersed beam. 
Thus, scanning the dispersed beam over the aperture causes the distribution first to enter and subsequently to leave the aperture, producing rising and falling edges in the measured current. 
Numerical differentiation of (filtered) $I(E)$ then produces positive and negative edge signals.
Provided that these signals are sufficiently separated, each provides an estimate of the beam energy distribution, allowing its width to be determined independently both sides of the aperture and detected spectra. 
The resulting width represents the combined response of the incident beam and the MEMS monochromator and should not be interpreted as the intrinsic energy resolution of either component alone. 
\subsection{Measured energy distributions}
\begin{figure}
    \centering
    \begin{subfigure}{0.49\textwidth}
        \includegraphics[width=\textwidth]{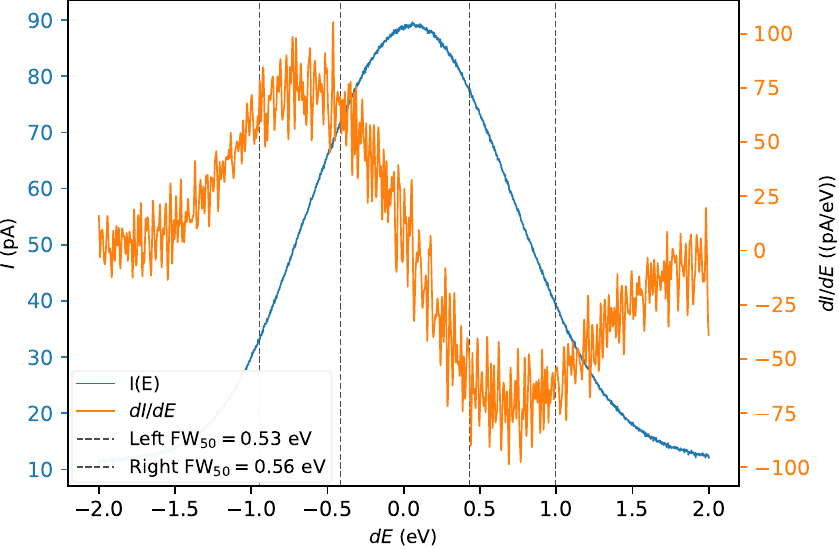}
        \caption{}
        \label{fig:sub1}
    \end{subfigure}
    \hfill
    \begin{subfigure}{0.49\textwidth}
        \includegraphics[width=\textwidth]{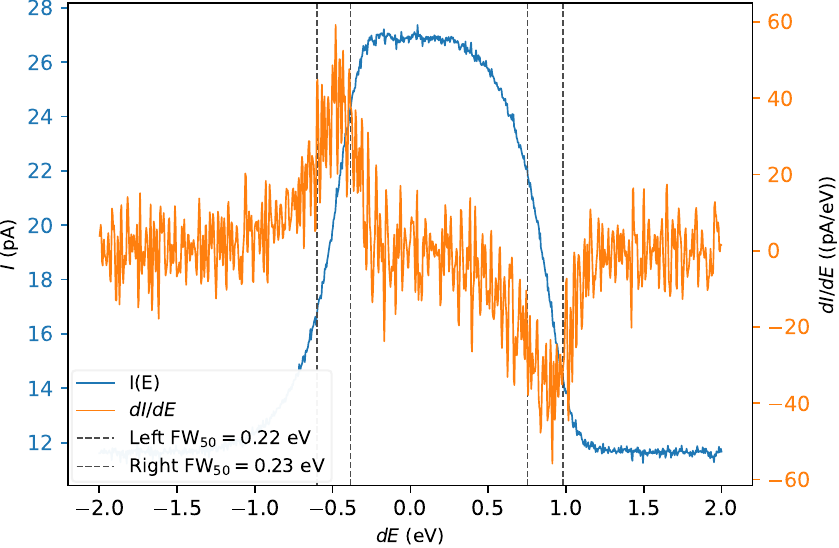}
        \caption{}
        \label{fig:sub2}
    \end{subfigure}
    \begin{subfigure}{0.49\textwidth}
        \includegraphics[width=\textwidth]{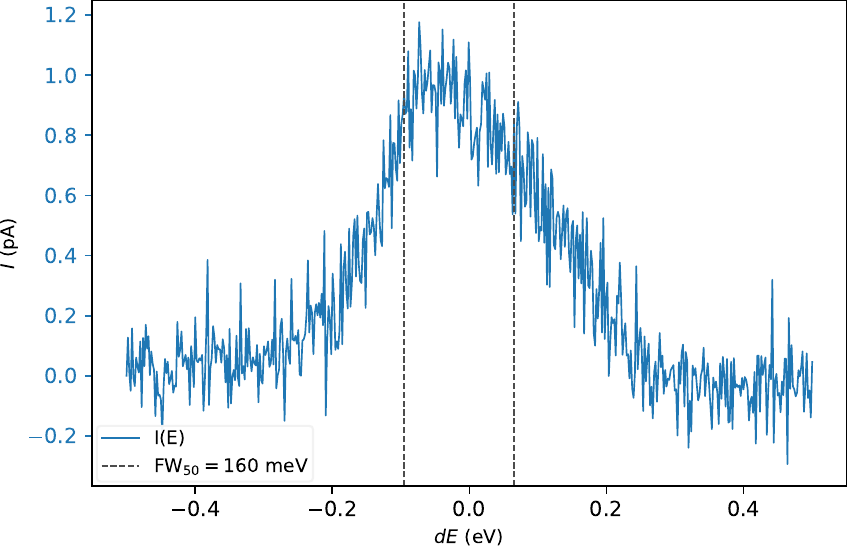}
        \caption{}
        \label{fig:sub3}
    \end{subfigure}

    \caption{Energy distributions measured using the MEMS monochromator. 
    (a) Reference measurement with the UC monochromator disabled and the beam transmitted through the coarse-alignment apertures. The transmitted current $I(E)$ and its numerically filtered derivative $\frac{\mathrm{d}I}{\mathrm{d}E}$ yield $\textrm{FW}_{50}$ values of 0.53 and 0.56 eV, respectively obtained from left and right sides of the curve. 
    (b) Corresponding measurement with the UC monochromator enabled, yielding $\textrm{FW}_{50}$ values of 0.22 and 0.23 eV. 
    (c) Direct transmission spectrum measured with the UC monochromator enabled and a 600 nm diameter MEMS aperture, yielding an $\textrm{FW}_{50}$ of 0.16 eV.
    }
    \label{fig: main }
\end{figure}

The measured transmission signals and their numerically filtered derivatives are shown in \autoref{fig: main }. The energy distributions are characterized using the full width containing half of the transmitted current ($\textrm{FW}_{50}$).

\autoref{fig:sub1} shows the reference measurement obtained with the UC monochromator disabled and the MEMS monochromator aligned to the coarse-alignment apertures. 
For an unfiltered Schottky electron source, a $\textrm{FW}_{50}$ between approximately 0.5 and 1 eV is expected.
The measured $\textrm{FW}_{50}$ values determined from the left and right derivative signals are 0.53 and 0.56 eV respectively, which is within the anticipated range.
However, as is visible in the figure, the left and right halves of the spectrum for the unfiltered beam overlap. 
Hence, the actual beam energy spread could be more than the measured value.

The measurement was repeated with the UC monochromator enabled, as shown in \autoref{fig:sub2}. The corresponding left and right FW50 values are 0.22 and 0.23 eV respectively. 
The lower value compared to the unfiltered beam measurement demonstrates that the MEMS monochromator can measure differences in the energy spread of the incoming beam and therefore confirms its basic energy-resolving functionality. 

A third measurement was performed with the UC monochromator enabled and the beam aligned to a 600 nm diameter aperture rather than to a coarse-alignment aperture. 
Because of the smaller aperture diameter, the rising and falling edge responses overlap and form a single peak in the measured transmission curve. 
The resulting spectrum, shown in \autoref{fig:sub3}, has a $\textrm{FW}_{50}$ of 0.16 eV.

The UC monochromator has a reported energy spread of approximately 0.15 eV, while the simulated energy resolution of the MEMS monochromator is 0.019 eV \cite{adriaans2026design} when using a 200-nm-diameter aperture. 
Assuming that its energy resolution scales linearly with aperture diameter, the expected resolution with a 600 nm aperture is approximately $\frac{600}{200} \times 0.019 = 0.057$ eV. 
If the broadening contributions from the UC monochromator and the MEMS monochromator are statistically independent and approximately Gaussian, combining these values in quadrature gives an expected measured width of approximately 0.16 eV in agreement with our measured value. Thus, this measurement confirms the energy-resolving capability of our MEMS monochromator. We note however, that the measurement does not independently demonstrate the predicted intrinsic resolution of 0.019 eV because the measured width is dominated by the energy spread of the incident beam produced by the UC monochromator.

\section{Discussion and conclusions}

We have demonstrated the fabrication and assembly of a miniaturized electrostatic MEMS-based electron monochromator. 
The device consists of nine electrode elements produced with five distinct chip designs. 
We used Bosch deep reactive-ion etching to fabricate the standard silicon electrode layers, while KOH etching and focused-ion-beam milling were used to produce the $\textrm{Si}_3 \textrm{N}_4$ membrane based nano apertures. 
The electrode chips were coated with molybdenum, passively aligned, bonded using borosilicate-glass spacers, and wire bonded to a flex PCB that provides the external electrical connections. 

The presented fabrication and assembly process is suitable for prototype production, but larger-volume manufacturing would require several modifications. Lithographically patterned and etched glass spacers could provide cleaner surfaces and more tightly controlled dimensions than laser-cut spacers. Likewise, replacing serial FIB milling with lithographic fabrication of the nano-apertures could enable parallel wafer-scale production. This would, however, require accurate alignment between the apertures of the different patterns on both sides of the wafer. 
Direct bonding could also eliminate the use of epoxy and reduce the amount of potentially outgassing material, but it would require improved control of surface flatness, cleanliness, and layer alignment, as well as a redesigned alignment procedure that does not rely on lateral adjustment after contact.

We provided a proof of principle experiment using the energy filtering capabilities of a MEMS monochromator prototype to measure the energy spread in an SEM in two different beam geometries. 
These measurements demonstrate the basic energy-resolving functionality of the MEMS monochromator. 
Smoother spectra could in principle have been obtained by increasing the number of sampled energy points and averaging repeated measurements. 
In practice, however, the beam energy was observed to drift by approximately 0.2 eV over acquisition periods of about 10 min, accompanied by a change in beam focus. 
An earlier retarding-field measurement conducted in our lab yielded an apparent energy spread of approximately 0.3 eV \cite{vos2021retarding}, which may have included a similar contribution from temporal beam-energy instability. 

Using the larger coarse alignment apertures, we have shown results with the UC monochromator both off and on. For the smaller 600 nm aperture, which yields a direct transmission measurement of the energy spectrum, we have only showed a result with the UC monochromator enabled. We have also attempted to repeat this measurement with the UC monochromator disabled. However, because of the larger energy spread of the unfiltered beam, chromatic aberration in the microscope optics prevented the complete energy distribution from being focused into AP0. 
The resulting energy-dependent transmission into the MEMS monochromator then truncates the incident spectrum preventing an accurate measurement of the full energy distribution.

The MEMS monochromator design has a theoretically and numerically predicted energy resolution of 19 meV \cite{adriaans2026design}, compared to a 160 meV proof-of-principle result presented here. Our measurement result could potentially be reduced by further tuning the voltages applied to the MEMS monochromator electrodes in order to sharpen the beam focus. The time available for this optimization was however limited because after prolonged exposure, the apertures started to clog, probably due to electron-beam-induced deposition of residual hydrocarbons in the microscope chamber. Further evaluation of the MEMS monochromator and assessment of its operational limits would thus benefit from a clean high vacuum environment and potentially the use of a glue-less fabrication scheme as discussed above.

Smoother spectra and more accurate measurements could also have been obtained using a higher input beam current. However, our SEM could deliver only about 15 pA at 500 eV electron energy. As a consequence, the full current capacity of approximately 11 nA could also not be evaluated using the present experimental configuration. In a dedicated experimental setup, monochromator placement would thus preferably be close to the electron source to minimize input brightness losses and allow for substantially higher input current. Combined with improved voltage stability and cleaner vacuum conditions, such a setup should permit a more direct evaluation of the energy resolution and current-handling capability of the MEMS monochromator.

In conclusion, we have presented the micro-fabrication of a miniaturized electron monochromator. As a basic proof of principle result, we measured a 0.16 eV $\textrm{FW}_{50}$ energy spread of a monochromated SEM beam  consistent with the expected combined response of both our MEMS and the SEM monochromator. 



\bibliographystyle{elsarticle-num} 
\bibliography{bibliography.bib}



\end{document}